\documentclass[10pt]{article}
\usepackage{epsfig}
\providecommand\bnabla{\mathbf{\nabla}}
\begin{document}
\centerline{\Large{TOPOGRAPHICAL SCATTERING OF WAVES: A SPECTRAL APPROACH}}
\vspace{1cm} \centerline{ by R. MAGNE$^1$, F. ARDHUIN$^2$, V. REY$^3$ and T. H.
C. HERBERS$^4$}\vspace{1cm} $^1$ Laboratoire de Sondages Electromagn\'etiques
de l'Environnement Terrestre, Universit{\'e} de Toulon et du Var, BP 132, 83957
La Garde cedex, France and Centre Militaire d'Oc{\'e}anographie, Service
Hydrographique et Oc{\'e}anographique
de la Marine, 13, rue du Chatellier 29609 Brest cedex, France\\
$^2$ Dr., Centre Militaire d'Oc{\'e}anographie, Service
Hydrographique et Oc{\'e}anographique de la Marine, 13, rue du
Chatellier 29609 Brest cedex, France\\
$^3$ Dr., Laboratoire de Sondages Electromagnétique de
l'Environnement Terrestre, Universit{\'e} de Toulon et du Var,
83957 La Garde cedex, France\\
$^4$ Prof., Dept. of Oceanography, Naval Postgraduate School,
Monterey, CA 93943, USA

\section*{Abstract}

The topographical scattering of gravity waves is investigated
using a spectral energy balance equation that accounts for first
order wave-bottom Bragg scattering. This model represents the
bottom topography and surface waves with spectra, and evaluates a
Bragg scattering source term that is theoretically valid for small
bottom and surface slopes and slowly varying spectral properties.
The robustness of the model is tested for a variety of
topographies uniform along one horizontal dimension including
nearly sinusoidal, linear ramp and step profiles. Results are
compared with reflections computed using an accurate method that
applies integral matching along vertical boundaries of a series of
steps. For small bottom amplitudes, the source term representation
yields accurate reflection estimates even for a localized
scatterer. This result is proved for small bottom amplitudes $h$
relative to the mean water depth $H$. Wave reflection by small
amplitude bottom topography thus depends primarily on the bottom
elevation variance at the Bragg resonance scales, and is
insensitive to the detailed shape of the bottom profile. Relative
errors in the energy reflection coefficient are found to be
typically $2h/H$.

\section*{\textit{CE DATABASE Subject Headings}}
Surface waves, scattering, wave reflection, spectral analysis.

\section*{INTRODUCTION}

Wave propagation over any bottom topography can now be predicted
with boundary element methods or other accurate numerical
techniques. However, wave forecasting relies to a large extent on
phase-averaged spectral wave models based on the energy or action
balance equation (Gelci et al. 1957) \nocite{gelci}. For large
bottom slopes waves can be reflected and this reflection is
currently not represented in these models, while the significance
of this process is still poorly known (Long 1973; Richter et al.
1976; Ardhuin et al. 2003).\nocite{Long}\nocite{Richteretal76} For
waves propagating over a sinusoidal seabed profile, a maximum
reflection or resonance is observed when the seabed wavenumber is
twice as large as the surface  wave wavenumber (Heathershaw
1982).\nocite{Heathershaw82}  Davies and Heathershaw (1984)
\nocite{D&H84} proposed a deterministic wave amplitude evolution
equation for normally incident waves over  a sinusoidal seabed,
based on a perturbation expansion for small bottom undulations.
This theory was shown to be in good agreement with experimental
data but overestimates reflection at resonance. Mei
(1985)\nocite{Mei85} developed a more accurate approximation that
is valid at resonance using a multiple scale theory. This approach
was further extended to random bottom topography in one dimension
(Mei and Hancock, 2003). The Bragg resonance theory can be
extended to any arbitrary topography in two dimensions, that is
statistically uniform(Hasselmann 1966).\nocite {Hasselman66}
Ardhuin and Herbers (2002) further included slow depth variations.
The resulting spectral energy balance equation contains a bottom
scattering source term $S_{\mathrm{bscat}}$, which is formally
valid for small surface and bottom slopes and slowly varying
spectral properties. $S_{\mathrm{bscat}}$  is readily introduced
into existing energy-balance-based spectral wave models, and was
numerically validated with field observations (Ardhuin et al.
2003).\nocite{Aetal03} While this stochastic theory is in a good
agreement with deterministic results for small amplitude
sinusoidal topography (Ardhuin and Herbers
2002)\nocite{Ardhuin2002}, the assumed slowly varying bottom
spectrum is not compatible with isolated bottom features, and the
limitations and robustness of the source term approximation for
realistic continental shelf topography are not well understood.
The limitations of the stochastic source term model are examined
here through comparisons with a deterministic model for arbitrary
one-dimensional ($1D$) seabed topography that is uniform along the
second horizontal dimension. We review the random Bragg scattering
model, and investigate the applicability limits of the source term
for a variety of seabed topography. Predicted reflection
coefficients are compared with results based on Rey's (1992)
model, which approximates the bottom profile as a series of steps.
Examples include modulated sinusoidal topography that is well
within the validity constraints of the source term approximation
as well as a steep ramp and a step that violate the assumption of
a slowly varying bottom spectrum and thus provide a simple test of
the robustness of the source term approximation.

\section*{THEORETICAL BACKGROUND}
\subsection*{MATCHING BOUNDARY SOLUTION}
We use Rey's (1992)\nocite{Rey92} algorithm, based on the theory
of Takano (1960)\nocite{Takano} and Kirby and Dalrymple
(1982)\nocite{D&K83}. It uses  a decomposition of the bottom
profile in a series of $N$ steps with integral matching along
vertical boundaries between  each pair of adjacent steps. A
coordinate frame is defined with the horizontal $x$ coordinate in
the direction  of the incident waves and the vertical $z$
coordinate pointing upwards relative to the mean water level. The
velocity potential is described by a sum of flat bottom
propagating and evanescent modes. Evanescent modes are included in
the matching condition to ensure a consistent treatment of the
wave field (Rey 1992). The general solution of the velocity
potential for a step ($p$) of depth $H_p$ is given by the
following equations:

\begin{equation}
\Phi_p(x,z,t)=\phi_p(x,z)\mathrm{e}^{-\mathrm{i}wt} ~~~~
\mbox{for} ~~~~ p=1,N , \label{laplace}
\end{equation}
with,
\begin{equation}
\phi_p(x,z)=\underbrace{A_p^{\pm}e^{\pm
ikx}\chi_p(z)}_{\mbox{propagating modes}}+
\underbrace{\sum_{q=1}^Q B_{p,q}^{\pm}e^{\pm k_q x}\psi_{p,q}(z)}_{\mbox{evanescent modes}%
},
\end{equation}
where $(\chi_p$, $\psi_{p,q} ~~~ q=1,Q)$ define a complete
orthogonal set for each step region ($p$):
\begin{equation}
\chi_p(z)=\cosh k_p(H_p+z) ,
\end{equation}
\begin{equation}
\psi_{p,q}(z)=\cos k_{p,q}(H_p+z).
\end{equation}
$k_p$ and $k_{p,q}$ satisfy the following dispersion relations:
\begin{equation}
\frac{\omega_p^2}{g}=k_p\tanh(k_pH_p),  \label{disprel}
\end{equation}
\begin{equation}
\frac{\omega_p^2}{g}=-k_{p,q}\tan(k_{p,q}H_p) .
\end{equation}
where g is the acceleration of gravity. \newline

\begin{figure}[h]
\centerline{\epsfig{file=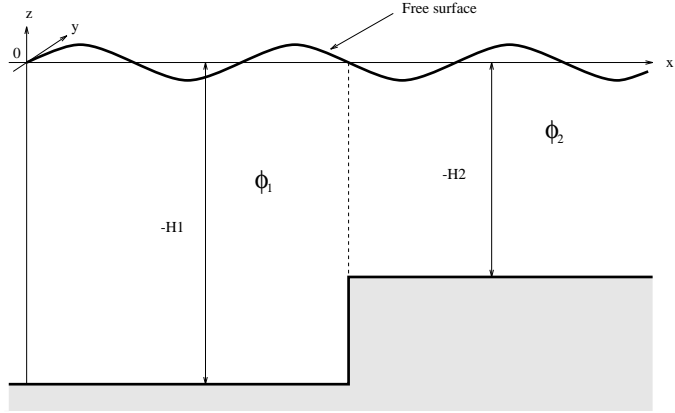,angle=-90,width=8.89cm}}
\caption{Stepwise approximation} \label{2marches}
\end{figure}
Across each step ($p$), matching conditions between two domains
(labelled $p=1$ and $p=2$ in figure \ref{2marches}) must be
applied to ensure continuity of the fluid velocity and surface
elevation.

\begin{equation}\label{}
  \phi_1=\phi_2,~ \frac{\partial{\phi_1}}{\partial x}=\frac{\partial{\phi_2}}{\partial
 x} ~~~~\mbox{for} ~~~~ -H_2<z<0,
\end{equation}
\begin{equation}\label{}
 \frac{\partial{\phi_1}}{\partial x}=0~~~~ \mbox{for} ~~~~
 -H_1<z<-H_2.
\end{equation}
The integral formulation of these conditions (for $H_1>H_2$) leads
to:
\begin{eqnarray}
\int_{0}^{H_2} \phi_1 \cdot \chi_2 dz & =& \int_{0}^{H_2} \phi_2
\cdot
\chi_2dz ,\\
\int_{0}^{H_2} \phi_1 \cdot \psi_{2,q} dz & =& \int_{0}^{H_2}
\phi_2 \cdot
\psi_{2,q}dz ~~~~\mbox{for} ~~~~ q=1,Q , \\
\int_{0}^{H_1} \frac{\partial{\phi_1}}{\partial x} \cdot \chi_1 dz
& = &
\int_{0}^{H_2} \frac{\partial{\phi_2}}{\partial x} \cdot \chi_1dz, \\
\int_{0}^{H_1} \frac{\partial{\phi_1}}{\partial x} \cdot
\psi_{1,q} dz & = & \int_{0}^{H_2}
\frac{\partial{\phi_2}}{\partial x} \cdot \psi_{1,q}dz
~~~~\mbox{for} ~~~~ q=1,Q .
\end{eqnarray}

The orthogonality of the set functions largely simplifies these
equations. In order to solve the problem numerically, the number
of evanescent modes $q$ are truncated to $q=Q$. Practically, only
a few evanescent modes are needed to ensure convergence.  For $N$
steps, $2N(Q+1)$ equations are solved to obtain the $2N(Q+1)$
complex coefficients $A_p^{\pm}$ and $B_{p,q}^{\pm}$. At the
boundaries (p=0 and p=N), the reflection coefficient is given by:
\begin{equation}
K_r=\frac{|A_0^-|}{|A_0^+|}
\end{equation}
\newline
This method has the advantage that it is valid for arbitrary $1D$
topography.

\subsection*{BRAGG SCATTERING THEORY}

We consider random waves propagating over a 2D irregular bottom
with a slowly varying mean depth $H$ and small-scale topography
$h$. The bottom elevation is given by $z=-H(\underline{\mathbf
x})+h(\underline{\mathbf x})$, with $\underline{\mathbf x}$ the
horizontal position vector. The free surface position is
$\zeta(\underline{\mathbf x},t)$.

\begin{figure}[h]
\centerline{\epsfig{file=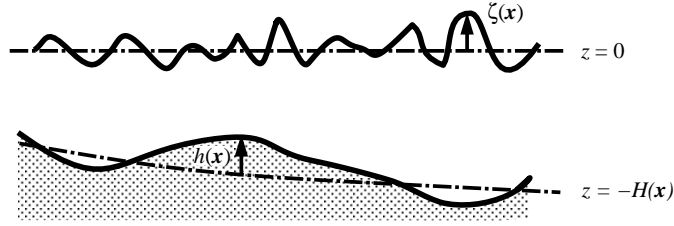,width=8.89cm}}
\caption{Definitions} \label{hetH}
\end{figure}

Considering an irrotational flow for an incompressible fluid, we
have the
governing equations and boundary conditions for the velocity potential $\phi$:\\
\begin{equation}
\nabla^2\phi+\frac{\partial^2{\phi}}{\partial{z^2}}=0 ~~~~
\mbox{for} ~~~~ -H+h \leq z\leq \zeta, \label{laplace}
\end{equation}
\begin{equation}
\frac{\partial{\phi}}{\partial z}=\underline{\nabla}\phi \cdot
\underline{\bnabla}(h-H) ~~~~ \mbox{at}~~~~ z=-H+h, \label{fond}
\end{equation}
\begin{equation}
\frac{\partial\zeta}{\partial t}=\frac{\partial{\phi}}{\partial z}
~~~~ \mbox{at} ~~~~ z=\zeta, \label{surflibre} \label{surf}
\end{equation}
\begin{equation}
g\zeta + \frac{\partial{\phi}}{\partial t}= -\frac{1}{2} [
|\underline{\bnabla} \phi|^2+(\frac{\partial{\phi}}{\partial z})^2
]
 ~~~~ \mbox{at} ~~~~ z=\zeta,
\label{Bernoulli}
\end{equation}
where $\underline{\bnabla}$ and $\nabla^2$ are the horizontal
gradient and Laplacian operators. The equations (\ref{laplace}),
(\ref{fond}), (\ref{surf}) and (\ref{Bernoulli}) are respectively
the Laplace's equation, free surface and bottom boundary
conditions, and
 Bernoulli's equation. Combining these two last equations, we obtain:
\begin{equation}
\frac{\partial^2{\phi}}{\partial{t^2}}+g\frac{\partial\phi}{\partial
z}= g\underline{\bnabla}\phi \cdot \underline{\bnabla}\zeta -
\underline{\bnabla}\phi \cdot
\frac{\partial{\underline{\bnabla}\phi}}{\partial t} -
\frac{\partial\phi}{\partial z}\frac{\partial^2\phi}{\partial
t\partial z} ~~~~ \mbox{at} ~~~~ z=\zeta. \label{comb}
\end{equation}
Assuming that the surface and the small-scale bottom slopes are of
the same order $\varepsilon$, and the large scale bottom slope is
of order $\varepsilon^2$, a perturbation expansion of $\phi$ up to
the third order in $\varepsilon$ yields the following spectral
energy balance equation (details are given in Ardhuin and Herbers
2002) :
\begin{equation}
\frac{dE(\underline{\mathbf{k}},\mathbf{x},t)}{dt}=
S_{\mathrm{bscat}(\underline{\mathbf{k}},\underline{\mathbf{x}},t)},
\label{consnrj1}
\end{equation}
where
\begin{equation}
S_{\mathrm{bscat}}\left( \underline{\mathbf{k}},\underline{\mathbf{x}},t\right)
=K\left( k,H\right) \int_{0}^{2\pi}\cos^{2}\left( \theta -\theta ^{\prime
}\right) F^{B}\left( \underline{\mathbf{k}}-\underline{\mathbf{k}^{\prime
}},\underline{\mathbf{x}}\right) \left[ E\left( \underline{\mathbf{k}^{\prime
}},\underline{\mathbf{x}},t \right) -E\left(
\underline{\mathbf{k}},\underline{\mathbf{x}},t \right) \right]
~\mathrm{d}\theta ^{\prime }, \label{S}
\end{equation}
with
\begin{equation}
K\left( k,H\right) =\frac{4\pi\omega k^{4}}{\sinh \left(
2kH\right) \left[ 2kH+\sinh \left( 2kH\right) \right] }.
\label{KofwH}
\end{equation}
$E(\underline{\mathbf{k}},\underline{\mathbf{x}},t)$ is the
surface elevation spectrum and
$F^B(\underline{\mathbf{k}},\underline{\mathbf{x}})$ is the
small-scale bottom elevation spectrum. These spectra are slowly
varying functions of $(\underline{\mathbf{x}},t)$ and
$\underline{\mathbf{x}}$ resectively. $\underline{\mathbf{k}}$ is
the wavenumber vector defined by
$\underline{\mathbf{k}}\equiv(k\cos\theta,k\sin\theta)\equiv
(k_x,k_y)$, where $\theta$ defines the angle with the $x$-axis.
 The spectral densities $E$ and $F^B$ are defined such that the integral over the entire $\underline{\mathbf{k}}$-plane
  equals the local variance,
\begin{equation}\label{}
  <h^2(\mathbf{\underline{x}})>=\int_{-\infty}^{+\infty}\int_{-\infty}^{+\infty}F^{B}(\underline{\mathbf{k}},\underline{\mathbf{x}})dk_xdk_y.
\end{equation}
 The frequency $\omega$ is given by the dispersion relation:
 \begin{equation}
\omega^2=gk\tanh(kH).
 \end{equation}
Here we consider a steady wave field in one dimension with
incident and reflected waves propagating along the x-axis. After
integration over $k_y$, $k_x$ becomes $k$ and (\ref{consnrj1})
reduces to
\begin{equation}
 C_g \frac{\partial E(k,x)}{\partial x}+C_k \frac{\partial E(k,x)}{\partial k} = S_{\mathrm{bscat}}(k,x),
\label{consnrj}
\end{equation}
\\
with a source term
\begin{equation}
S_{\mathrm{bscat}}(k,x)=K(h,H)\frac{F^B(2k,x)}{k} \left[ E(-k,x)-E(k,x)
\right]. \label{I}
\end{equation}
The first term of Eq.(\ref{consnrj}) represents advection in
physical space with the group velocity defined by
\begin{equation}
C_g=\frac{dx}{dt}=\frac{\partial\omega}{\partial k},
\end{equation}
and the second term describes the effect of shoaling on the
wavenumber
\begin{equation}
C_k=\frac{dk}{dt} = -C_g \cdot \frac{2k^2}{2kh+\sinh(2kh)} \cdot
\frac{\partial H}{\partial x}\cdot
\end{equation}

\section*{REFLECTION BY MODULATED SINUSOIDAL BOTTOM TOPOGRAPHY}

The source term approximation was validated by Ardhuin and Herbers
(2002)
for random waves reflecting from a sinusoidal seabed, by integrating $S_{%
\mathrm{bscat}}$ analytically across the wave spectrum in the
limit of weak reflection ($E(-k)<<E(k)$ , with positive and
negative wavenumbers corresponding to the incident and reflected
waves, respectively). A comparison with Dalrymple and Kirby's
(1986) \nocite{D&K86} solution gave good agreement, even for only
a few bars. For stronger reflection, equation (\ref{consnrj}) is
not readily evaluated analytically, and numerical integration is
not feasible since a sinusoidal bottom has an infinitely narrow
spectrum (a Dirac distribution), and thus cannot be represented
with a finite bottom discretization $\Delta k_b $.

We consider instead a bottom spectrum with a finite width that
corresponds to a modulated sinusoidal bottom profile.

\begin{figure}[h]
      \centerline{\epsfig{file=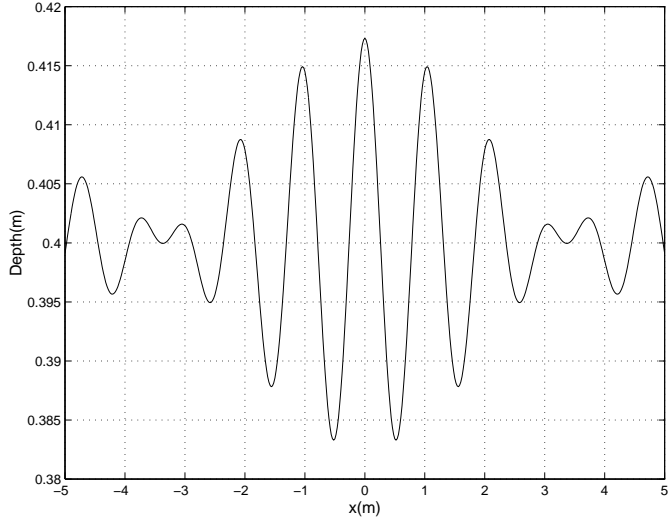,width=8.89cm}}
      \caption{Modulated seabed (m=3), $bk_{b,0}=0.06$}
      \label{fdmod}
\end{figure}
\begin{figure}
      \centerline{\epsfig{file=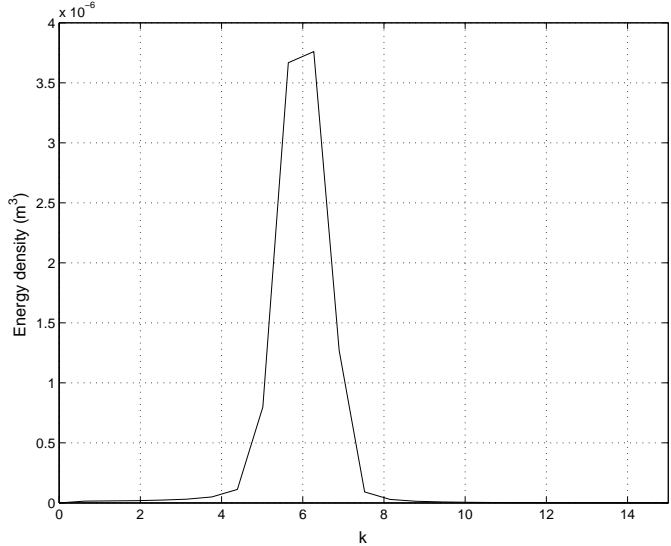,width=8.89cm}}
      \caption{Modulated seabed spectrum (m=3)}
      \label{Spfdmod}
\end{figure}
\begin{figure}[h]
\centerline{\epsfig{file=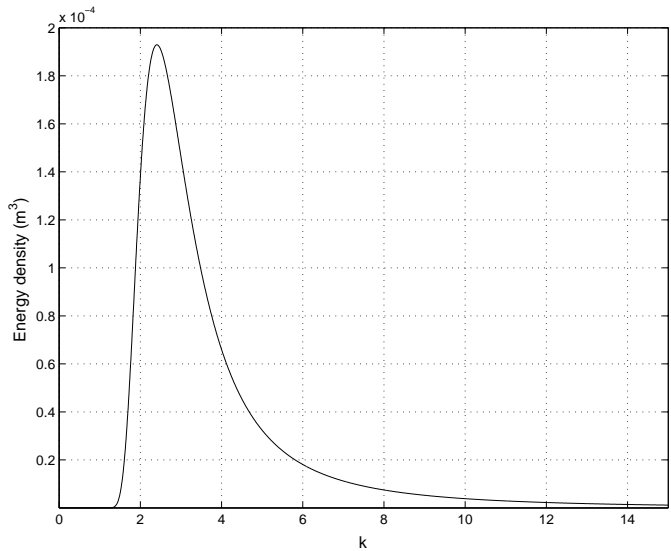,width=8.89cm}} \caption{Incident
wave spectrum} \label{Spwave}
\end{figure}
\begin{figure}[h]
\centerline{\epsfig{file=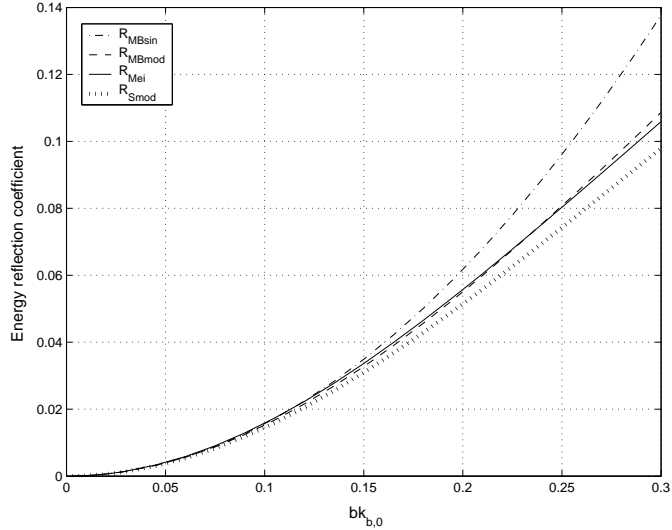,width=8.89cm}} \caption{Wave
reflection by modulated sinusoidal bottom} \label{Ermod}
\end{figure}

The modulated seabed is represented by a sum of cosines:
\begin{equation}
h(x)=\Sigma^{i=(m-1)/2}_{i=-(m-1)/2} b_i cos[(k_{b,0}+i\Delta
k_b)x]
\end{equation}
The slowly varying depth ($H$), defined in part $2$ is taken
constant while the perturbation ($h$) represents the modulated
seabed. We define the root mean square (r.m.s.) bar amplitude b
from the bottom
variance, $b=\sqrt{<h^2>}$, and a representative bottom slope $%
\varepsilon=bk_{b,0}$. The reflected wave energy is calculated for
the bed profile shown in figure \ref {fdmod}, with the peak bottom
wave number $k_{b,0}=6 \mathrm{m}^{-1}$
($\lambda_{b,0}=1.04\mathrm{m}$), and a short modulation length
with $m=3$, and equal amplitudes $(b_i)$ for all bottom
components. The length of the bed is 1.5 modulation lengths,
giving the bottom spectrum shown in figure \ref{Spfdmod}. The
reflection from this modulated sinusoidal bottom was evaluated for
an incident Pierson-Moskowitz spectrum, with a peak at $k_0$
satisfying the Bragg resonance condition $2k_0=k_{b,0}$
(Fig.\ref{Spwave}). Spectral results for Rey's model were obtained
by evaluating reflection coefficients for monochromatic waves over
a range of frequencies and integrating the reflected energy across
the spectrum. $70$ steps are used to resolved the bathymetry.
Results for various values of $b$ are displayed in the form of
reflection coefficients $R$ (Fig.\ref{Ermod}) as a function of the slope $%
bk_{b,0}$. $R$ is defined by the ratio of the reflected and
incident energies: $R={\ ( \Sigma_{k<0}E ) }/{\ ( \Sigma_{k>0}E )
}$. Predictions based on the source term method
($R_\mathrm{Smod}$) and the
 matching boundary model using 5 evanescent modes ($R_{\mathrm{MBmod}}$) agree well over a wide range of
bottom slopes. The solutions gradually diverge for large bottom
slopes where the source term underpredicts the reflection. Even
for $bk_0=0.3$ ($b/\lambda_0=0.05$), differences are less than $10
\%$ confirming the robustness of the source term method for steep
topography.

To evaluate the effect of the spectral width on the reflection
coefficient, figure \ref{Ermod} also includes predictions for
sinusoidal topography ($m$=0) with the same variance. Results for
sinusoidal topography were obtained using Mei's (1985) analytical
approximation and Rey's (1992) algorithm.
The resulting reflection coefficients $R_{\mathrm{Mei}}$ and $%
R_{\mathrm{MBsin}}$, respectively, agree for small bottom slopes
(Fig.\ref {Ermod}) and diverge for larger slopes as already shown
by Rey (1992). Indeed, $R_{\mathrm{Mei}}$ was derived for small
bottom slopes while the matched boundary solution converges to the
exact reflection for any bottom profile when the number of
evanescent modes goes to infinity. What may seem surprising is
that the reflection coefficient for the sinusoidal and modulated
sinusoidal topographies $R_{\mathrm{MBmod}}$ and
$R_{\mathrm{MBsin}}$ agree for small slopes although bottom
profiles are quite different. Apparently, for small bottom slopes
and narrow bottom spectra the reflection is only a function of the
total bottom elevation variance $b^2$ and does not depend on the
phases of its components. This result is obvious from the
viewpoint of the source term theory that was derived for small
bottom slopes, and does not retain the phases of the bottom
spectrum components. The predicted reflection depends on the
convolution of the wave spectrum with the bottom spectrum at the
Bragg resonance wavenumber (the integral of (\ref{I}) over all
wavenumbers). If the bottom spectrum is narrow compared with the
wave spectrum then the total source term depends only on the total
bottom variance and the surface spectral density at the Bragg
resonance wavenumber.

\section*{REFLECTION BY A LINEAR RAMP}
\begin{figure}[h]
\centerline{\epsfig{file=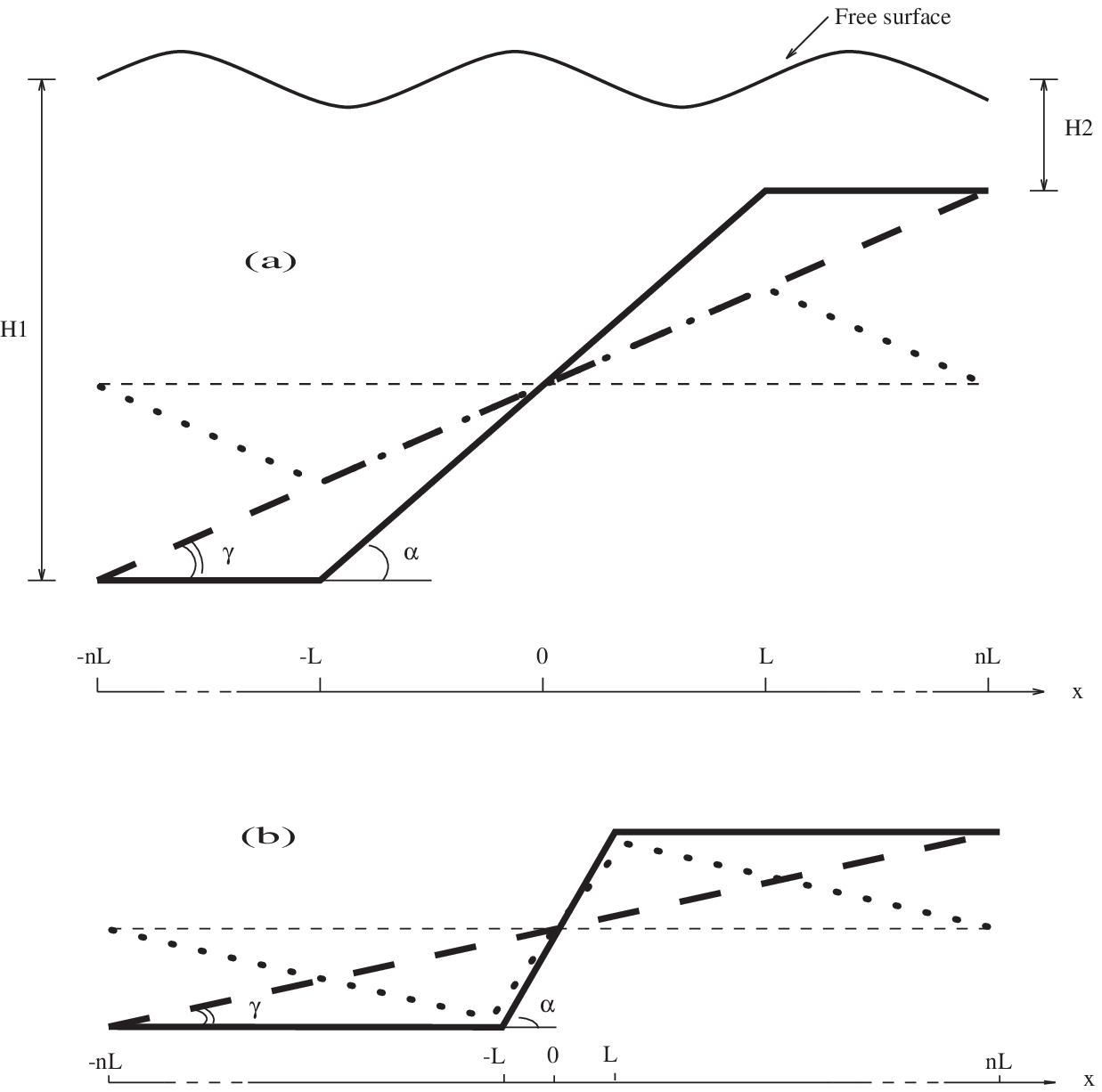,angle=0,width=8.89cm}} \caption{Decomposition
of linear ramp (solid line) into a slowly varying depth $H$ (dashed line) and
residual $h$ (dotted line). (a) and (b) for small and large $n$ respectively}
\label{ramp}
\end{figure}

To investigate the robustness of the variance-based source term
model for reflection induced by localized topography, we consider
the linear ramp problem used in previous studies to test the mild
slope equation (Booij 1983)\nocite{Booij}. In the source term
approximation, wave scattering is the result of interactions
between surface waves and bottom variations at the scale of the
surface wavelength. The scattering model is thus based on a
decomposition of the topography into a slowly varying depth $H$
and a perturbation $h$ (small scale topography), which corresponds
to a separation between refraction and shoaling that occurs over
the slowly varying depth $H$ and scattering at these short scales.
For practical applications, it is desirable to have a perturbation
$h$ that is zero outside of a finite region, so that the spectrum
of $h$ is well defined. Once the two criteria that the slope of
$H$ does not exceed a given threshold and $h$ is zero outside of a
region of radius $nL$ are satisfied, the choice of the depth
decomposition in $h$ and $H$ is fairly arbitrary and does not
affect the following results.
For simplicity we take a piecewise linear function for $H(x)$, so
that the perturbation $h(x)$ takes the form of a triangular wave
(Fig.\ref{ramp}).

The ramp profile is defined by the fixed water depths $H_1$,
$H_2$, while the ramp slope $\alpha$ is varied by adjusting its
length $2L$ (Fig.\ref{ramp}). To ensure that $H(x)$ is slowly
varying, $\gamma$ has to be small. This is achieved by extending
the domain to a length $2nL$ with $n>1$ (Fig.\ref{ramp}). The
slope of H is then given by $\tan\ \gamma ={(\tan\ \alpha)}/n$,
with several values of $n$ tested below.

\subsection*{First test case: small depth change}

\begin{figure}[h]
\centerline{\epsfig{file=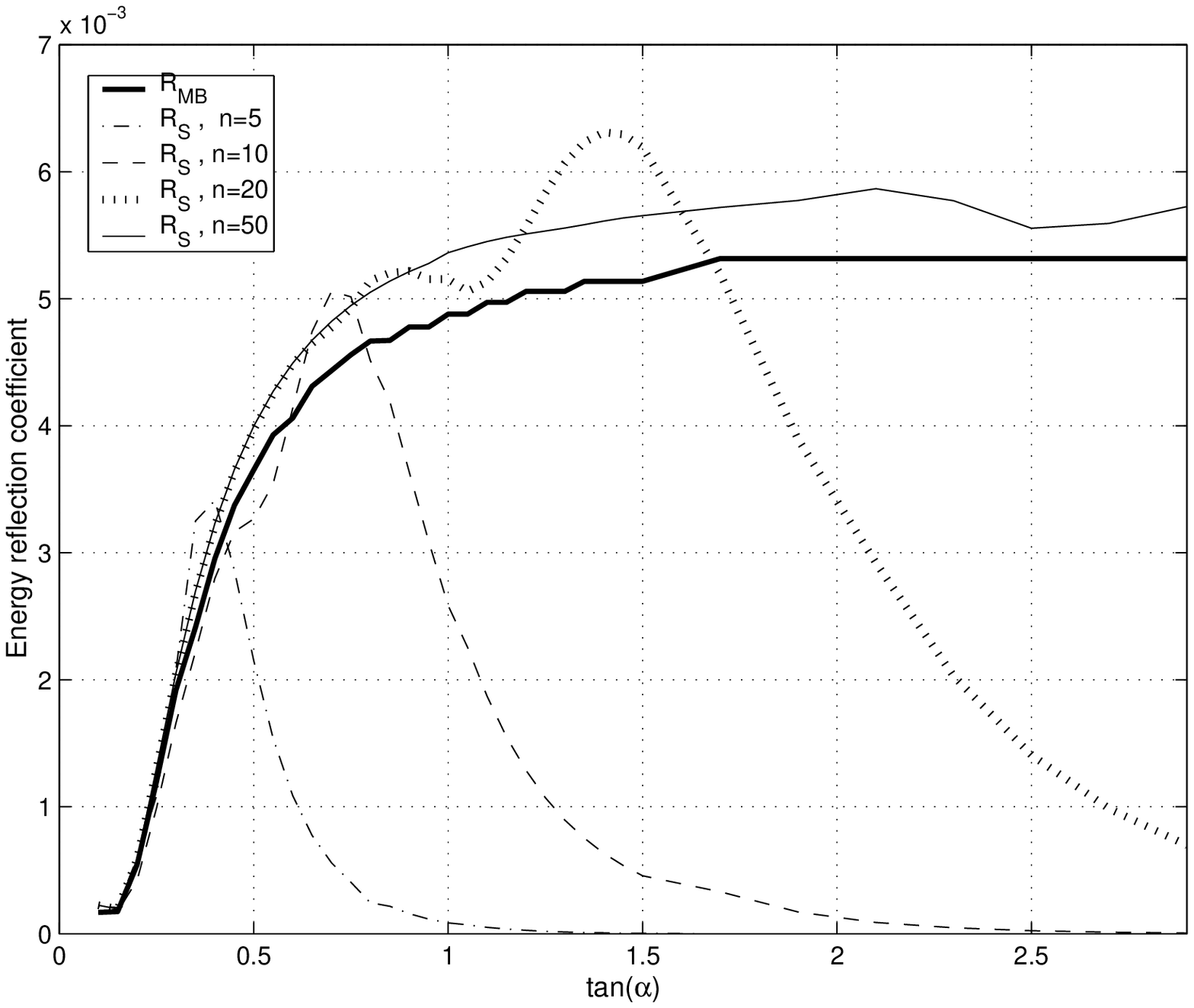,width=8.89cm}} \caption{Wave
reflection by a ramp} \label{Eramp}
\end{figure}

\begin{figure}[h]
\centerline{\epsfig{file=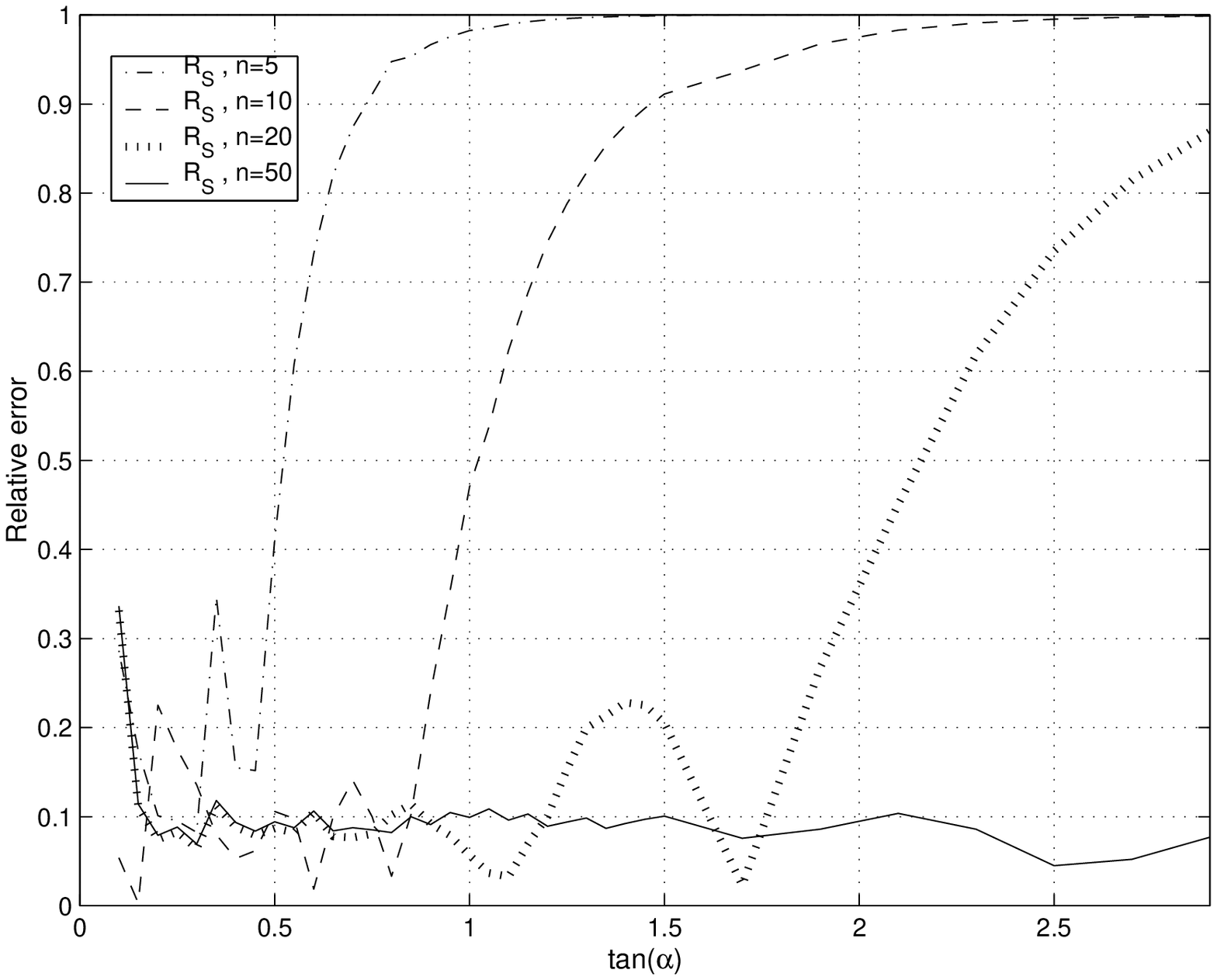,width=8.89cm}} \caption{Relative
errors in wave reflection by a ramp} \label{Err}
\end{figure}

We first consider a ramp with a small depth transition from $H_1=0.5\mathrm{m}$ to $%
H_2=0.3\mathrm{m}$. The incident wave spectrum is represented by
the same Pierson-Moskowitz spectrum that was used in the previous
section with the peak wavenumber in deep water
$k_0=3\mathrm{m}^{-1}$ (Fig.\ref{Spwave}), so that $k_0H_1=1.5$
and $k_0H_2=0.9$. In order to
investigate the source term applicability limits, the linear ramp slope $%
\tan\alpha$ is varied from $0.01$ to $2.9$. For each value of
$\alpha$, several values of $\gamma$ are tested, with $n$ varying
from $5$ to $50$. The reflection coefficient $R_\mathrm{S}$
(source term reflection due to the residual) is compared with the
''exact'' computation $R_{\mathrm{MB}}$
(matching boundary algorithm) in figure \ref{Eramp} and the relative error $%
(R_\mathrm{S}-R_{\mathrm{MB}})/R_{\mathrm{MB}}$ is shown in figure
\ref{Err}. In our calculations, for slopes of $H$ such as
$\tan\alpha < 0.4,\ R_{\mathrm{S},n=5}$ is within $30\%$ of the
exact value $R_{\mathrm{MB}}$. For
larger values of $\tan\alpha$, $R_{\mathrm{S},n=5}$ decreases and tends to zero (Fig.%
\ref{Eramp}), while the exact solution $R_{\mathrm{MB}}$ converges
to the reflection over a vertical step as $\tan\alpha$ goes to
infinity. The value $\tan\alpha=0.4$ corresponds to $\tan
\gamma~(={\tan\alpha}/5)$ equal to $0.08$. For larger $n $ the
slope of $H$ is reduced and $R_{\mathrm{S},n}$ is valid for a
wider range of ramp slopes.

We notice that for all values of $n$ shown in figure \ref{Eramp},
the model gives reasonable results for $\tan \gamma~ (=
(\tan\alpha)/n)$ up to about $0.08$. The ramp slope does not
appear to be a limiting factor (as it was assumed in the theory).
For $\tan \gamma$ larger than $0.08$ the reflection is
increasingly underestimated probably because of the contribution
of the large scale profile $H(x)$ to the reflection.
As $n$ increases $h$ approaches the slope of the actual ramp and
$R_{\mathrm{S},n}$ converges to $R_{\mathrm{S},\infty}$ which is
about $10\%$ larger than $R_{\mathrm{MB}}$ for all ramp slopes. As
discussed below, the accuracy of the model is apparently not
limited by the ramp slope.

It may seem surprising that $R_{\mathrm{S},n}$ actually converges
for large $n$ while the bottom spectrum does not. In the case of a
vertical step of height $h$ in the middle of a domain of length
$2nL$, the spectral density $F^B(k)$ of a discrete variance
spectrum of the residual is proportional to $h^2 /2nLk^2$ and
tends to zero (except around $k = 0$) as $n$ goes to infinity.
However the source term formulation represents scattering as
uniformly distributed along the bottom, and the integration of the
source term along the wave propagation path yields a reflection
that is proportional to $2nL \ F^B(k)$ and thus converges when $n$
goes to infinity.
 The use of infinite support for $H$ and $h$
(taking the limit $n\rightarrow\infty$) to compute the reflection
over a localized ramp is counterintuitive. It represents a
physically localized scattering with a mathematically distributed
source. In practice, the bottom spectrum is obtained by discrete
Fourier Transform of the bottom, and it only tends to continuous
power spectrum in the limit $n\rightarrow\infty$. Further, it
should be realized that the bottom power spectrum is the Fourier
transform of the bottom autocorellation function used by Mei and
Hancock (2003, see Appendix).

For a non-random bottom such as the ramp here, one may use intermediate results
by Mei and Hancock (2003)\nocite{Mei&Hancock2003} where the hypothesis that the
bottom is random only comes in for discarding nonlinear wave effects (which are
not taken into account here). It thus appears that our rather surprising result
for the convergence as $n\rightarrow\infty$ is justified by the convergence of
the discrete spectrum to the continuous power spectrum and the theory of Mei
and Hancock (2003) applied to non-random bottoms (see Appendix). It shows that
the far field scattered energy by small amplitude depth variations only depends
on the power spectrum of the scatterers at the Bragg scale, and not on its
localization in space, as long as the bottom amplitude remains small.

\subsection*{Booij's ramp: larger depth change}
\begin{figure}[h]
\centerline{\epsfig{file=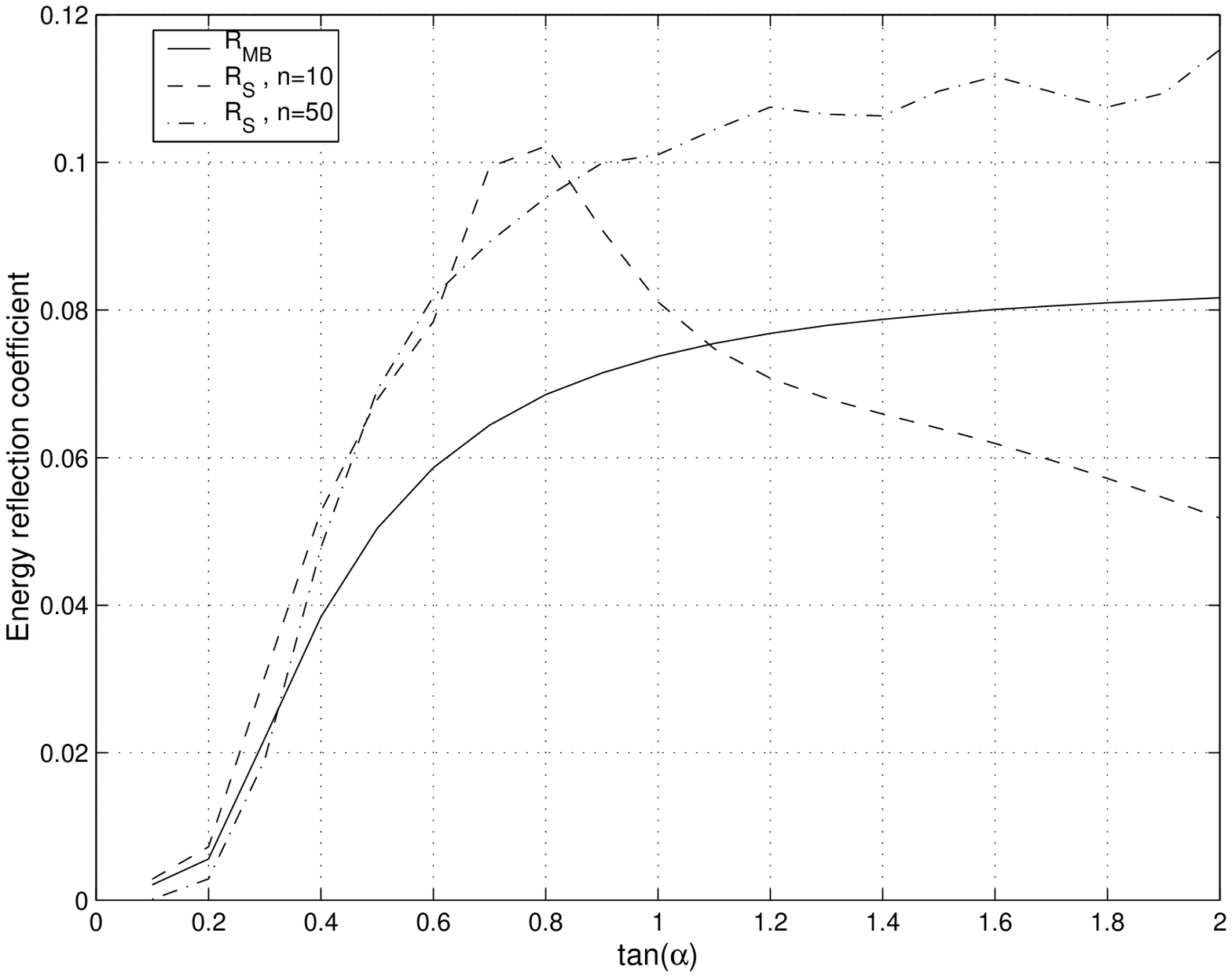,width=8.89cm}} \caption{Wave
reflection by Booij's ramp} \label{ERrampebooij}
\end{figure}

This approach should clearly break down for finite bottom amplitudes, in
particular because sub-harmonic scattering was observed (Belzons et al. 1991)
while it is not explained by the present theory. Such a limit should be tested
to see whether our present approach has some practical applicability. We
therefore take a second test case with a larger ramp is taken from Booij (1983)
with water depths $H_1=4.97$m, $H_2=14.92$m and an incident wave peak period
$T=10$s. The corresponding peak wavenumber in deep water
$k_0=0.04\mathrm{m}^{-1}$ so that $k_0H_1=0.6$ and $k_0H_2=0.2$. Results for
ramp slopes $\tan \alpha$ ranging from 0.001 to 2.9, and n=10 and 50 are shown
in figure \ref {ERrampebooij}.

We notice again that $R_{\mathrm{S},n}$ converges for large n,
provided that $(\tan\alpha )/n\ <\ 0.08$. However, in this case
the relative error is larger than in the first test case, up to
about $30\%$. The two tests have the same ramp slopes but
different ratio of water depths at the edges of the ramp
$H_1/H_2=3$ here versus $H_1/H_2=1.7$ in the previous case. The
two cases suggest
 that the source term is more sensitive to the amplitude
than the slope of the bottom perturbation $h$. Formally, the
bottom amplitude only appears in the bottom boundary condition
(\ref{fond}), which is linearized at $z=-H$ using the following
Taylor series expansion;
\begin{equation}
\Phi |_{z=-H+h}=\Phi |_{z=-H}+h\frac{\partial \Phi }{\partial z}|_{z=-H}+%
\frac{h^{2}}{2}\frac{\partial ^{2}\Phi }{\partial
z^{2}}|_{z=-H}+O(h^{3}). \label{bob}
\end{equation}
Ardhuin and Herbers(2002) use a representative length scale
$1/k_{0}$ to non-dimensionalize (\ref{bob}) as,

\begin{equation}
\tilde{\Phi}|_{\tilde{z}=-H+h}=\tilde{\Phi}|_{\tilde{z}=-H}+\eta\frac{%
\partial \tilde{\Phi}}{\partial \tilde{z}}|_{\tilde{z}=-H}+\frac{%
\eta^{2}}{2}\frac{\partial ^{2}\tilde{\Phi}}{\partial \tilde{z}^{2}}%
|_{\tilde{z}=-H}+O(\eta^3),  \label{taylorexp}
\end{equation}
where $\tilde{z}=k_0z$, $\eta=k_0h$, $\eta$ corresponding to the
scales that cause wave scattering. The validity of the Taylor
expansion requires that $\eta$ is small and also that the first
and second derivative of $\tilde{\phi}$ with respect to
$\tilde{z}$ are of order $1$. In this approximation
(\ref{taylorexp}) is limited by the small-scale
 slope $k_0h$. However one may also take $H_0$ as the
representative length which leads to the same equation
(\ref{taylorexp})
 with $\eta=h/H_0$, limited then by the water depth ratio $h/H_0$.
 The choice of the representative length was arbitrary and can be justified only a posteriori, by evaluating the scale
  of variation of $\Phi$ and thus the magnitude of $\partial \tilde{\Phi}/\partial \tilde{z}$ and
  $\partial ^{2}\tilde{\Phi}/\partial \tilde{z}^{2}$.
 The numerical results presented here show that the
source term is more sensitive to the water depth change $h/H_0$
than the small-scale slope $k_0h$. Booij (1983) had found that the
standard mild slope equation (Berkhoff 1972\nocite{Berkhoff1972})
gave errors less than $10\%$ for $\tan \alpha$ up to $1/3$. Our
results suggest that the Bragg scattering model can be as accurate
as the mild slope equation for computing reflection, but only for
$\Delta h/H_0$ less than $0.2$.


\section*{REFLECTION BY A STEP}
\begin{figure}[h]
\centerline{\epsfig{file=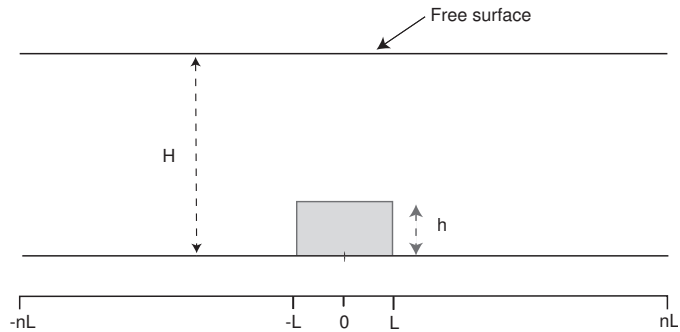,width=8.89cm}} \caption{Sketch
of the step} \label{Step}
\end{figure}

Now that the effect of $h/H_0$ is well established, one may
question the importance of other parameters. We thus evaluate
source term predictions of broad and narrow surface wave spectra
over steps of varying height to gain further insight into the
 limitations of the source term approximation for localized
topography. Reflection of waves by a rectangular step has been
investigated analytically and experimentally in numerous studies
(Neuman 1965a,b\nocite{Neuman65a}\nocite{Neuman65b}; Miles
1967\nocite{Miles67}; Mei and Black 1969\nocite{Mei&Black69}; Mei
1983\nocite{Meibook} and Rey, Belzons and Guazzelli
1992)\nocite{ReyJFM92} and is well understood. The step is defined
in figure \ref{Step}, where $2L$ is the step-length, $h$ the
height and $2nL$ the size of the entire computational domain.

\subsection*{Numerical set-up}
\begin{figure}[h]
\centerline{\epsfig{file=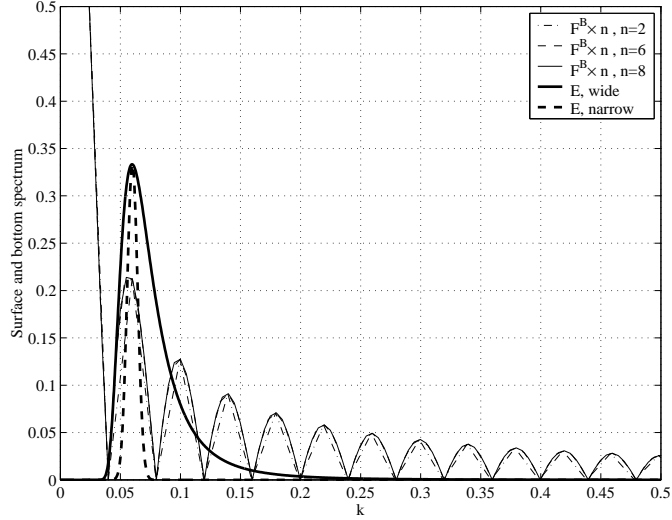,width=8.89cm}} \caption{Wide
and narrow surface wave spectra superposed on the bottom spectrum
for domain sizes $n=2,6\mbox{ and }8$. The bottom spectrum is
rescaled by the surface wavenumber ($F^B(k/2)$) to show the
resonant bottom and surface components.} \label{Spf-vs-Spw}
\end{figure}

The spectral density of the bottom $F^B(k)$ is proportional to
$h^2/2nLk^2$. Hence, integration of the source term along the wave
propagation path yields a reflection that is proportional to
$2nLF^B(k) \sim h^2/k^2$, independent of $n$. Although the domain
length has no effect on real waves in the absence of bottom
friction, it influences the discretization of the bottom spectrum
($\Delta k=2\pi/2nL$), and thus it may have an impact on the
numerical results. However $2nF^B(n)$ converges as $n$ goes to
infinity (Fig.\ref{Spf-vs-Spw}), so that the domain length does
not change the results for large enough values of $n$. A large
domain with $n=8$ was used here.

The step width ($2L$) is taken to be half the wavelength of the
surface waves for a spectrum peak $k_{0p}=0.04$m$^{-1}$
($L_0=157\mathrm{m}$) in a water depth of $15\mathrm{m}$. Two
different wave spectra are used here (bold lines in figure
\ref{Spf-vs-Spw}): a wide spectrum (solid) with a classic
Pierson-Moskowitz shape, typical of wind seas, and the narrow
swell-like spectrum (dashed) with a Gaussian shape.
 Once the shape of wave spectrum is chosen, the solution is a function of three non-dimensional variables:
the step height $h/H$, the water depth $k_{0p}H$, and the relative
step width $k_{0p}L$.

\subsection*{Influence of the height of the step}

\begin{figure}[h]
\centerline{\epsfig{file=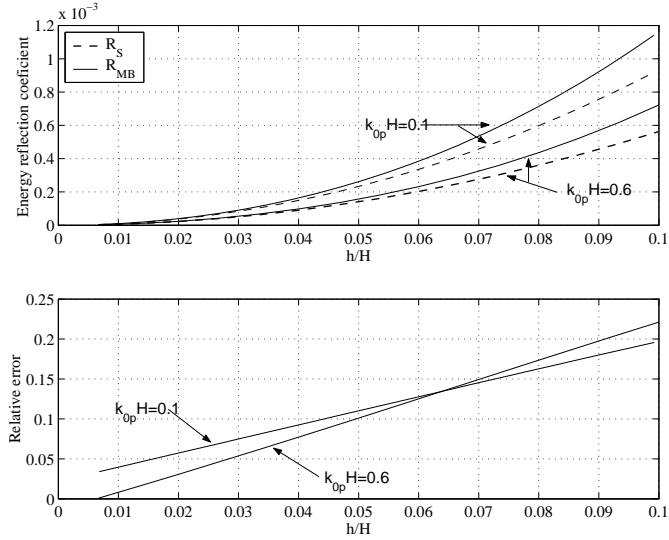,width=8.89cm}}
\caption{Reflected energy computed with the source term
(dash-line) and with the matching boundary algorithm (full-line),
for intermediate depth ($k_{0p}H=0.6$) and shallow water
($k_{0p}H=0.1$), and relative error of the source term.}
\label{Kr_kh1kh2}
\end{figure}

The accuracy of the source term for a range of non-dimensional
step heights $h/H$ is evaluated in intermediate and shallow water
through comparison with the "exact" matching boundary algorithm
(Fig.\ref{Kr_kh1kh2}). Energy reflection coefficients are compared
for two different water depths, $k_{0p}H=0.1$ and $k_{0p}H=0.6$,
representative of shallow and intermediate depths. The incident
wave spectrum has a Pierson-Moskowitz shape.

As expected from previous calculations, the error in the source
term increases with the step amplitude $h/H$. For $h/H<0.05$ the
error in the predicted reflection coefficients is less than
$10\%$. These results provide further confirmation that the height
of the localized scatterer is a limiting factor for the source
term computation, but not its slope, which is infinite here, and
this result holds for very shallow water.

\subsection*{Influence of the width of the step and the wave spectrum}

\begin{figure}[h]
\centerline{\epsfig{file=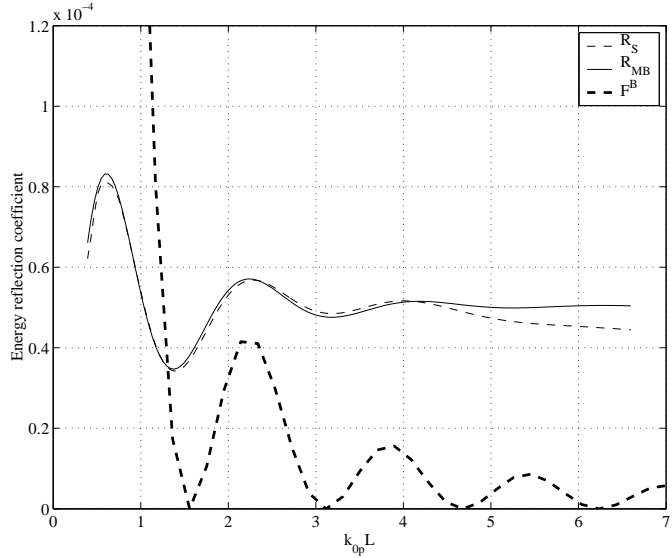,width=8.89cm}}
\caption{Reflected energy computed with the source term
(dash-line) and with the matching boundary algorithm (full-line)
for a wide wave spectrum. The bottom spectrum ($F^B$) is also
indicated, scaled by the normalized resonant surface wavenumber to
indicate the resonant response (bold dash-line). Other parameters
are $h/H=0.02$ and $k_{0p}H=0.1$.} \label{BarKr_koL_wide}
\end{figure}

\begin{figure}[h]
\centerline{\epsfig{file=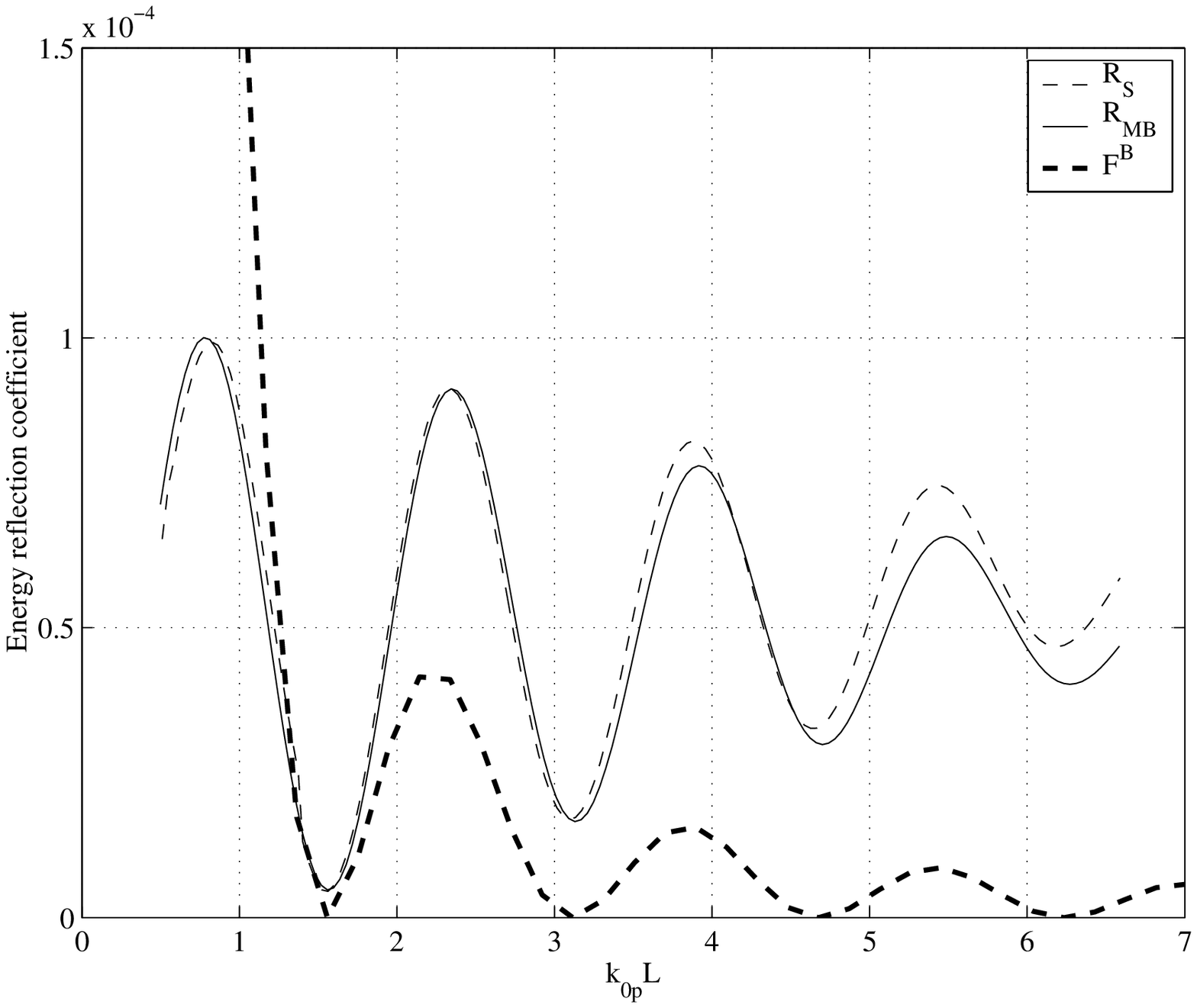,width=8.89cm}} \caption{Same as
Fig.\ref{BarKr_koL_wide} but for a narrow wave spectrum. }
\label{BarKr_koL_narrow}
\end{figure}

Here we consider the dependence of the reflection coefficient on
the width of the step and the width of the wave spectrum
 for a small amplitude step ($h/H=0.02$) in shallow water ($k_{0p}H=0.1$). The
non-dimensional step width $k_{0p}L$ is varied, effectively
changing the position of the wave spectrum peak  relative to the
bottom spectral peaks (see Fig.\ref{Spf-vs-Spw}). Results are
shown in figures \ref{BarKr_koL_wide} and \ref{BarKr_koL_narrow}
for wide and narrow wave spectra, respectively.

The same computation is done for the narrow spectrum
(Fig.\ref{BarKr_koL_narrow}).

For both wide and narrow surface waves spectra, the source term
yields accurate results,
 and the errors do not appear to be sensitive to the width of the step. Oscillations
 in the reflection coefficient with varying $k_{p0}L$ represent an interference phenomenon
 that has been described in numerous previous studies. When a
monochromatic incident wave runs up the leading edge of the step
at $x=-L$, it is partly reflected and partly transmitted. As the
transmitted component passes the rear edge of the step at $x=L$,
it is again partially reflected and partially transmitted. If the
reflected waves originating from the front and rear edges of the
step are in phase we have a constructive interference which
amplifies the reflection. Conversely, destructive interference
occurs if the two reflected wave trains are $180$ degrees out of
phase and cancel out, yielding zero reflection. For long waves,
maximum reflection occurs when $\sin^2 2k_{p0}L=1$ (Mei, 1983),
where
$k_{p0}$ is the incident wave wave number. This condition is met when:\\
\begin{equation}\label{maxrefl}
    2k_{0p}L=(2n-1)\frac{\pi}{2},\qquad n=1,2,3,\cdots
\end{equation}
The corresponding values of $k_{p0}L$ are $k_{p0}L = \pi/4 \simeq
0.78,\ 3\pi/4 \simeq 2.35,\ 5\pi/4 \simeq 3.93 \cdots$ These
values match with the reflection peaks observed in the figures
(\ref{BarKr_koL_wide},\ref{BarKr_koL_narrow}) both for the source
term and the matching boundary algorithm. In the wide spectrum
case (Fig.\ref{BarKr_koL_wide}) these oscillations are suppressed
and for high values of $k_{p0}L$, the reflection tends to a
constant value. Using Bragg scattering, this is explained by the
fact that in the limit of large step width $k_{p0}L$ the wave
spectrum is wider than the side lobes of the bottom spectrum (see
Fig.\ref{Spf-vs-Spw}) and the effects of constructive and
destructive interferences for different spectral component average
out. The reflection coefficient is a convolution of the bottom
spectrum and the surface wave spectrum, and thus the reflection is
insensitive to bottom spectral details with scales finer than the
wave spectrum width.

\section*{CONCLUSIONS}

Predictions of the scattering of surface waves by bottom
topography based on a spectral energy balance equation that
includes a wave-bottom Bragg scattering source term (Ardhuin and
Herbers 2002) are compared with exact results based on a matching
boundary algorithm (Rey 1992). The source term yields accurate
reflection predictions for modulated sinusoidal topography. In the
limit of small bottom amplitudes $h$ compared to the water depth
$H$, the two models yield identical results, confirming that the
far-field scattered wave is determined entirely by the variance
spectrum of the bottom and does not depend on the phases of its
components. This finding also holds for localized topography, a
result that can be justified by the approach of Mei and Hancock
(2003) using their intermediate results for non-random bottoms. In
that case, the bottom spectrum must be carefully calculated over a
large enough domain in order to resolve the important bottom
scales. Using Discrete Fourier Transforms, one may use an
artificial gently sloping extension of the area covered by
scatterers. However, it is found that it also holds for very steep
topography, such as a single step, for a variety of water depths
and wave spectrum shapes, as long as $h<H$ is small. In our
calculations, relative errors in the energy reflection coefficient
are found to be typically $2h/H$, or $h/H$ for the amplitude
reflection coefficient. These results show that the Bragg
scattering source term is a reasonably accurate method for
representing wave reflection in spectral wave models, for a wide
range of small amplitude bottom topographies found on continental
shelves. The source term approach is also very efficient compared
to the elliptic models such as proposed by Athanassoulis and
Belibassakis (1999).\nocite{Athanassoulis1999} An extension of the
source term to higher order (e.g. following Liu and Yue
1998)\nocite{LiuandYue98} may reduce errors for larger values of
$h/H$, that are shown here to be the limiting factor in practical
applications. Results for $1D$ bottom profiles are expected to
hold for practical $2D$ applications of the source term
approximation.

\section*{Acknowledgements}

This research is supported by a joint grant from Centre National
de la Recherche Scientifique (CNRS) and D\'el\'{e}gation
G\'en\'erale pour l'Armement (DGA). Additional funding is provided
by the U.S. Office of Naval Research, and the U.S. National
Science Foundation in the framework of the 2003 Nearshore Canyon
Experiment (NCEX). Fruitful discussions with Kostas Belibassakis
are gratefully acknowledged.

\section*{APPENDIX. RECONCILIATION OF RANDOM AND DETERMINISTIC WAVE THEORIES}
Mei and Hancock (2003) considered the same problem of a wave train propagating
over an arbitrary topography of small amplitude $h$. In their scaling $h$ is
small compared to the wavelength $2\pi/k$, but, as discussed in this paper, the
scaling for the bottom perturbation could also be the mean water depth $H$.
These authors further assume that $h$ is a random function that is stationary
with respect to the fast coordinate $x$, and introduce a slow coordinate $x_1$
for variations in the statistics of $h$. This two-scale approach is similar to
that used by Ardhuin and Herbers (2002). Mei and Hancock (2003) obtained an
amplitude evolution equation in which the topography acts as a linear damping
with a coefficient $\beta_i$ given by their equation (B8) as
\begin{equation}\label{beta}
\beta_i=\frac{\omega(k\sigma)^2k(\hat{\gamma}(2k)+\hat{\gamma}(0))}{4\cosh^2kH(\omega^2H/g+\sinh^2kH)},
\end{equation}
where $\sigma^2(x_1) \gamma$ is the auto-correlation function of
the bottom topography, decomposed in a slowly-varying local
variance $\sigma^2(x_1)$ and a normalized auto-covariance
$\gamma$. $\hat{\gamma}$ is the Fourier Transform of $\gamma$.
Although Mei and Hancock's (2003) result does not conserve energy
(which requires the introduction of higher order terms, see
Ardhuin and Herbers 2002), it is rather general as far as the
bottom is concerned. The essential difference with Ardhuin and
Herbers (2002) is that there is no need for a large number of
bottom undulations to obtain an expression for the scattering, and
the "number of undulations" is properly defined by the scale over
which the auto-covariance goes to zero.

Naturally the two theories are consistent, and we can obtain from
$\beta_i$ the damping coefficient $\beta_E$ for the energy, which
is twice that for the wave amplitude $A$ since $\partial {(A
A^\star)}/\partial t= -2 \beta_i A A^\star = -\beta_E A A^\star$,
with $A^\star$ the complex conjugate of $A$. Re-writing
(\ref{beta}) one has,
\begin{equation}
\beta_E=\frac{2k^3\omega\sigma^2(\hat{\gamma}(2k)+\hat{\gamma}(0))}{\sinh2kH\left[2kH+\sinh2kH\right]}.
\end{equation}
For a zero-mean stationary process the Fourier transform of the
auto-covariance function is simply $2 \pi$ times the power
spectral density $F^B$ (e. g. Priestley
1981\nocite{Priestley1981}, theorem 4.8.1 p 211), so that, for
$F^B(0)=0$, we get
\begin{equation}
\beta_E= \frac{4 \pi k^3\omega
F^B(2k)}{\sinh2kH\left[2kH+\sinh2kH\right]},
\end{equation}
which is the linear part of the bottom scattering source term
(\ref{I}) in one dimension,
\begin{equation}
S_{\mathrm{bscat}} (k) =  \beta_E \left(E(-k)-E(k)\right).
\end{equation}

Interestingly the hypothesis of randomness for $h$ is not
important for the value of $\beta_i$ when averaged over the entire
field of scatterers (however, it does impact the real part, i.e.
the phase of the waves). Following Mei and Hancock's (2003)
derivation, one may define a $\beta_i$ that is also a function of
the fast coordinate $x$ using their equation (2.36), and in that
case the derivation is identical, replacing $\sigma^2(x_1) \gamma$
by $h(x) h(x-\xi)$, all the way to their equations (B1)--(B3).
Then one may define a mean value, which, in the case of a finite
region with scatterers between $-nL$ and $nL$ reads,
\begin{equation}
\overline{\beta_i}=\frac{1}{2nL}\int_{-nL}^{nL} \beta_i(x)
{\mathrm d}x.
\end{equation}

Taking the imaginary part of their equations (B1)--(B3) we have,
\begin{equation}
\overline{\beta_i}=\omega \frac{k^2}{2
\cosh^2(kH)}\left\{\frac{I_0}{\omega^2 H/ g + \sinh^2 (kH) }+
\sum_{n=1}^{\infty} \frac{k I_n}{k_n \left[\omega^2 H/ g + \sin^2
(k_n H)\right] }\right\}
\end{equation}
with (correcting a few minor type-setting errors in their paper),
\begin{equation}\label{I0}
I_0=-\Re\left\{\frac{1}{2nL}\int_{-nL}^{+nL}\int_{-\infty}^{+\infty}
\left(\frac{\mathrm d^2}{{\mathrm d}\xi^2}-{\mathrm i}k
\right)\left(h(x) h(x-\xi)\right) \mathrm{e}^{{\mathrm i} k \xi +
ik\left|\xi\right|} {\mathrm d}\xi {\mathrm d}x\right\}
\end{equation}
and
\begin{equation}\label{In}
I_n=-\Im\left\{\frac{1}{2nL}\int_{-nL}^{+nL}\int_{-\infty}^{+\infty}
\left(\frac{\mathrm d^2}{{\mathrm d}\xi^2} -{\mathrm i}k
\right)\left(h(x) h(x-\xi)\right) \mathrm{e}^{{\mathrm i} k \xi +
ik_n \left|\xi\right|} {\mathrm d}\xi {\mathrm d}x\right\}
\end{equation}
Switching the order of the integrals, \ref{I0}-\ref{In} are
identical to their equations (B2)--(B3), provided that we redefine
$\gamma$ as the full auto-covariance function
\begin{equation}
\gamma(\xi)=\frac{1}{2nL}\int_{-nL}^{+nL}h(x) h(x-\xi) {\mathrm
d}x.
\end{equation}
In this case $\gamma$ is obviously real and even and we obtain
their equation (B8) for $\overline{\beta_i}$.

We have thus proved that in one dimension and in the limit of
small bottom amplitudes the scattering source term applies to
non-random bottoms. In these conditions, the linear part of the
source term represents the damping of the incident waves (and thus
also the average scattered wave energy) averaged over the area
covered by scatterers.

\newpage

\section*{APPENDIX.B SYMBOLS}

$A$ = propagating modes amplitude;\\
$b$ = root mean square amplitude from the bottom variance;\\
$B$ = evanescent modes amplitude;\\
$Cg$ = group velocity;\\
$Ck$ = spectral advection velocity;\\
$E$ = surface elevation spectral density;\\
$F^B$ = small-scale bottom elevation spectrum;\\
$h$ = bottom perturbation height;\\
$H$ = water depth;\\
$k$ = surface wavenumber;\\
$k_0$ = peak wavenumber in deep water;\\
$k_{0p}$ = peak wavenumber;\\
$k_{b,0}$ = peak bottom wavenumber;\\
$K$ = Source term coefficient;\\
$Kr$ = amplitude reflection coefficient;\\
$L$ = half-length of the ramp;\\
$L_0$ = peak wavelength;\\
$n$ = mild slope inclination parameter;\\
$m$ = modulation parameter;\\
$R$ = energy reflection coefficient;\\
$R_{MB}$ = Matching Boundary energy reflection coefficient;\\
$R_{Mei}$ = Mei energy reflection coefficient;\\
$R_S$ = Source term energy reflection coefficient;\\
$S_{scat}$ = bottom scattering source term for the wave energy spectrum;\\
$T_0$ = peak period;\\
$\alpha$ = ramp inclination;\\
$\gamma$ = mild slope inclination;\\
$\varepsilon$ = representative bottom slope;\\
$\zeta$ = free surface position;\\
$\eta$ = small parameter;\\
$\Phi$ = velocity potential;\\
$\chi$, $\psi_n$ = complete orthogonal set of functions;\\
$\omega$ = wave radian frequency;\\
$\Delta
k_b$ = discretization of the bottom spectrum;\\

\subsection*{Subscripts}
$\tilde{}$ = non-dimensionalized variable;\\

\end{document}